\documentclass[]{elsarticle}
\usepackage{verbatim}
\usepackage{amssymb,amsfonts}
\usepackage{amsmath}
\usepackage{breqn}
\usepackage{lipsum, mathtools, cuted, breqn}
\usepackage{braket}
\usepackage{hyperref}

\makeatletter
\def\ps@pprintTitle{%
	\let\@oddhead\@empty
	\let\@evenhead\@empty
	\def\@oddfoot{}%
	\let\@evenfoot\@oddfoot}
\makeatother

\begin{document}

\begin{frontmatter}

\title{Influence of polarization and self-polarization charges on impurity binding energy in spherical quantum dot with parabolic confinement}

\author{Samrat Sarkar}

\author[]{Supratik Sarkar\corref{mycorrespondingauthor}}
\cortext[mycorrespondingauthor]{Corresponding author}
\ead{supratik.sarkar.1995@gmail.com}

\author[]{Chayanika Bose}

\address{Department of Electronics and Telecommunication Engineering, Jadavpur University, Kolkata 700032, India}

\begin{abstract}
We present a general formulation of the ground state binding energy of a shallow hydrogenic impurity in spherical quantum dot with parabolic confinement, considering the effects of polarization and self energy. The variational approach within the effective mass approximation is employed here. The binding energy of an on-center impurity is computed for a $\mathrm{GaAs}$/ $\mathrm{{Al}_{x}{Ga}_{1-x}As}$ quantum dot as a function of the dot size with the dot barrier as parameter. The influence of polarization and self energy are also treated separately. Results indicate that the binding energy increases due to the presence of polarization charge, while decreases due to the self energy of the carrier. An overall enhancement in impurity binding energy, especially for small dots is noted.
\end{abstract}


\end{frontmatter}

\section{Introduction}
\label{Intro}

As impurities in semiconductor nanostructures play a crucial role in determining the electronic and optical properties of quantum devices, study of impurity states in nano-scale attracted attention in the field of low-dimensional semiconductor research. With the pioneering work of Bastard \cite{bib1,bib2} in the early 80's, a lot of work on impurity states in quantum wells (QWs) \cite{bib3,bib4}, quantum well wires (QWWs) \cite{bib5,bib6} and quantum dots (QDs) \cite{bib7,bib8} has been reported in the last three decades. As the spatial confinement of carriers is stronger in QDs than in QWs and QWWs with the same confining dimensions, the bound states are most pronounced for impurities within a QD. Thus, the study of impurity binding energy in QDs becomes of prime interest. Theoretical studies involve QD models of a finite or infinite (square or parabolic) confining potential with off-center or on-center impurity. Perturbation method \cite{bib9,bib10,bib11} and variational method \cite{bib12,bib13,bib14,bib15} emerged as the main tool for the theoretical estimations. However in a finite-barrier QD, dielectric mismatch at the dot-barrier interface causes image charges formed by the ionized impurities and carriers. Presence of these charges provides an additional potential energy to the carriers that should further influence the impurity states. This aspect was investigated in refs. \cite{bib16,bib17,bib18}. Later, Movilla \textsl{et al.} \cite{bib11} estimated binding energies considering polarization and self-polarization energies for a spherical QD with a finite square well potential barrier. The same problem with second order band non-parabolicity was solved by Bose \textsl{et al.} \cite{bib20} using variational method. However, the binding energy for a shallow hydrogenic impurity in spherical QD with a finite parabolic potential well, taking both the polarization and self-polarization charges into consideration is yet to be studied. The problem will therefore, be addressed in the present communication. 

\par The variational technique will be employed to determine the ground donor state for the above QD. In the first section, the model for theoretical formulation will be developed with a shallow hydrogenic donor at an arbitrary location within the dot. Results computed for a $\mathrm{GaAs}$/ $\mathrm{Al_x{Ga}_{1-x}As}$ QD with an on-center donor will be presented in the next section. The ground donor state will be estimated for dots of different barrier heights and radii in order to find the influence of polarization and self-energy on it. The concluding remarks will be presented in the final section.


\section{Theory}
\label{Theory}

In the effective mass approximation, the envelop function $\psi$ of an electron in a finite barrier QD is described by the Schr{\"o}dinger equation as

\begin{equation}
\label{eq1}
\left[\frac{p^2}{2m^*}+V\left(r\right)\right]\psi = E\psi
\end{equation} 

\noindent where $p$, $m^*$ and $E$ are the momentum, effective mass and energy (w.r.t. the conduction band edge) of the electron. $V\left(r\right)$ is the perturbation term arising from the discontinuity in conduction band edge at the interface of the dot and the embedding materials.

\par If we assume parabolic confinement within the spherical QD, which provides a more realistic picture, $V\left(r\right)$ can be defined as

\begin{subequations}
	\label{eq2}
	\begin{equation}
	V\left( r \right) =  \frac{1}{2} m_w {\omega}^2 r^2 \quad \text{for $r\leq R$}
	\end{equation}
	\begin{equation}
	V\left( r \right) =  \frac{1}{2} m_w {\omega}^2 R^2 = V_0 \quad \text{for $r> R$} 
	\end{equation}
\end{subequations}

\noindent where $m_w$ is effective mass of electron in the well material, $\omega$ is frequency of the parabolic potential and $R$ is the dot radius. The electron wave function comes as

\begin{subequations}
	\label{eq3}
	\begin{equation}
	\psi\left(r\right)=A {\mathrm{exp}}\left(-\frac{\lambda r^2}{2}\right) {}_1^1 \mathrm{F} \left\lbrace \frac{1}{2} \left( \frac{3}{2} - \mu \right);\frac{3}{2};\lambda r^2 \right\rbrace \quad \text{for $r\leq R$}
	\end{equation}
	\begin{equation}
	\psi\left(r\right) = B \frac{{\mathrm{exp}}\left(-\chi r \right)}{r} \quad \text{for $r> R$}
	\end{equation}
\end{subequations}

\noindent where $A$ and $B$ are the normalized probability amplitudes of the wavefunction after taking into the boundary conditions, ${}_1^1 \mathrm{F}$ stands for confluent hypergeometric function, $\lambda = m_w^* \omega/ \hbar$, $\mu = E/ \hbar \omega$, $\chi = \sqrt{2m_b^* \left( V_0 - E \right)/ {\hbar}^2}$ and $m_b^*$ is effective mass of electron in the barrier material.

\par Applying the continuity condition of the wave functions and their space derivatives (Ben Daniel-Duke condition) at the dot boundary, we get the following equation 

\begin{equation}
\label{eq4}
\frac{{}_1^1 \mathrm{F} \left\lbrace \frac{1}{2} \left( \frac{3}{2} - \mu \right);\frac{3}{2};\lambda R^2 \right\rbrace}{{}_1^1 \mathrm{F} \left\lbrace \frac{1}{2} \left( \frac{7}{2} - \mu \right);\frac{5}{2};\lambda R^2 \right\rbrace}
=
\frac{1-\frac{2}{3} \mu}{1-\frac{m_w^*}{m_b^*}.\frac{\left(1+\chi R \right)}{\lambda R^2}}
\end{equation}

\noindent This is to be solved numerically to find out the ground state energy ($E_0$) of an electron confined in the dot.

\par In the presence of a shallow hydrogenic impurity located at any arbitrary position $r'$ from the centre of the QD, Eq. \ref{eq1} is modified as

\begin{equation}
\label{eq5}
\left[ \frac{p^2}{2m^*} + V\left(r\right) + V_c\left(r,r'\right) + V_p\left(r,r'\right) + V_{se}\left(r\right) \right] \psi = E \psi
\end{equation}

\noindent Here, $V_c\left(r,r'\right)$ is the coulomb interaction term between the donor and electron, and is described as

\begin{subequations}
	\label{eq6}
	\begin{equation}
	V_c \left(r,r'\right) = - \frac{e^2}{4 \pi {\epsilon}_0 {\epsilon}_w \left|r-r'\right|} \quad  \text{for $r\leq R$}
	\end{equation}
	\begin{equation}
	V_c \left(r,r'\right) = - \frac{e^2}{4 \pi {\epsilon}_0 {\epsilon}_b \left|r-r'\right|} \quad  \text{for $r> R$}
	\end{equation}
\end{subequations}

Effective energy of the electron due to the polarization charges induced by the impurity is included as $V_p\left(r,r'\right)$ in Eq. \ref{eq5}. The expression of polarization energy is obtained by solving the Poisson's equation for a charge $+e$ embedded in a sphere of static dielectric constant ${\epsilon}_w$ at the position $r'$ from its center, with the sphere itself surrounded by a material of dielectric constant ${\epsilon}_b$, and is given by \cite{bib21}

\begin{subequations}
	\label{eq7}
	\begin{equation}
	V_p\left(r,r'\right) = 
	-\frac{e^2}{4 \pi {\epsilon}_0 {\epsilon}_w R} \sum_{n=0}^{\infty} \frac{\left(n+1\right)\left({\epsilon}_w - {\epsilon}_b\right)}{n{\epsilon}_w+\left(n+1\right){\epsilon}_b} {\left(\frac{rr'}{R^2}\right)}^n P_n\left(cos \theta\right)  \quad  \text{for $r\leq R$}  
	\end{equation}
	\begin{equation}
	V_p\left(r,r'\right) = 
	-\frac{e^2}{4 \pi {\epsilon}_0 {\epsilon}_b r} \sum_{n=0}^{\infty} \frac{\left(n+1\right)\left({\epsilon}_w - {\epsilon}_b\right)}{n{\epsilon}_w+\left(n+1\right){\epsilon}_b} {\left(\frac{r'}{r}\right)}^n P_n\left(cos \theta\right) \quad  \text{for $r> R$}
	\end{equation}
\end{subequations}

\noindent where $\theta$ is the angle between the two vectors $r$ and $r'$, and $P_n$ is the $n^{th}$ Legendre polynomial.

\par The carrier itself induces polarization charges and the net potential energy of the carrier due to those charges, i.e., the self energy of the carrier is given by \cite{bib21}

\begin{equation}
\label{eq8}
V_{se} \left( r \right) = - \frac{1}{2} e {\phi}_{ind} \left( r \right)
\end{equation}

\noindent where ${\phi}_{ind} \left( r \right)$ is the electrostatic potential due to the self polarization charges. It can be obtained first by solving Poisson's equation for the potential at point $r$ due to the electron (charge $-e$) located at a position $r'$ from the dot center, and then for the limit $r' \rightarrow r$ with both $r'$ and $r$ located either in the well material or in the barrier material simultaneously. Finally $V_{se} \left( r \right)$ can be expressed as \cite{bib21}

\begin{subequations}
	\label{eq9}
	\begin{equation}
	V_{se}\left(r\right) = 
	-\frac{e^2}{8 \pi {\epsilon}_0 {\epsilon}_w R} \sum_{n=0}^{\infty} \frac{\left(n+1\right)\left({\epsilon}_w - {\epsilon}_b\right)}{n{\epsilon}_w+\left(n+1\right){\epsilon}_b} {\left(\frac{r}{R}\right)}^{2n}   \quad  \text{for $r\leq R$}  
	\end{equation}
	\begin{equation}
	V_{se}\left(r\right) = 
	-\frac{e^2}{8 \pi {\epsilon}_0 {\epsilon}_b r} \sum_{n=0}^{\infty} \frac{\left(n+1\right)\left({\epsilon}_w - {\epsilon}_b\right)}{n{\epsilon}_w+\left(n+1\right){\epsilon}_b} {\left(\frac{R}{r}\right)}^{2n+1} \quad  \text{for $r> R$}
	\end{equation}
\end{subequations}

In principle, the ground state binding energy for the donor located at any arbitrary position $r'$ from center of the QD, can be estimated by variational method. The trial wave function is taken as

\begin{subequations}
	\label{eq10}
	\begin{equation}
		\psi \left( r \right) = 
		N\left( \beta \right) \mathrm{exp}\left( - \frac{\lambda r^2}{2} \right)  {}_1^1\mathrm{F}\left\lbrace \frac{1}{2} \left( \frac{3}{2} - \mu \right); \frac{3}{2}; \lambda r^2 \right\rbrace 
		 \mathrm{exp} \left( - \beta \left| r-r' \right| \right)
		\quad \text{for $r \leq R$}
	\end{equation}
	\begin{multline}
		\psi \left( r \right) = 
		N\left( \beta \right) \mathrm{exp}\left( - \frac{\lambda R^2}{2} \right)  {}_1^1\mathrm{F}\left\lbrace \frac{1}{2} \left( \frac{3}{2} - \mu \right); \frac{3}{2}; \lambda R^2 \right\rbrace
		 R  \frac{\mathrm{exp}\left(\chi \left(R-r\right)\right)}{r} 
		 \\ \mathrm{exp} \left( - \beta \left| r-r' \right| \right)
		\quad \quad \text{for $r > R$}
	\end{multline} 
\end{subequations}

For the sake of simplicity we will calculate the binding energies for on-center impurity, i.e. $r' = 0$. If $H$ is the Hamiltonian of the carrier for $r' = 0$, then the energy of the carrier is given by

\begin{multline}
\label{eq11}
E_0^' \left( \beta \right) 
= 
{\bra{\psi\left(r\right)}}_{r'=0}  {H}  {\ket{\psi\left(r\right)}}_{r'=0} 
\\
= 4\pi N^2 \left( \beta \right) \left[ - \frac{{\hbar}^2}{2m_w^*} \left( \frac{3}{2} - \mu \right) \right. \int_0^R \left[ \mathrm{exp} \left( - \lambda r^2 \right) \mathrm{exp} \left( -2 \beta r \right) \left[ \left( 1 - \beta r - \lambda r^2 \right) \right.  \right.
\\
{}_1^1\mathrm{F}\left\lbrace \frac{1}{2} \left( \frac{3}{2} - \mu \right); \frac{3}{2}; \lambda r^2 \right\rbrace  + \frac{2}{3} \lambda r^2 \left( \frac{7}{2} - \mu \right) {}_1^1\mathrm{F}\left\lbrace \frac{1}{2} \left( \frac{11}{2} - \mu \right); \frac{5}{2}; \lambda r^2 \right\rbrace  - \frac{2}{3} \lambda r^2 
\\ 
\left. \left\lbrace \left( \frac{3}{2} - \mu \right) + \beta r + \lambda r^2 \right\rbrace {}_1^1\mathrm{F} \left\lbrace \frac{1}{2} \left( \frac{7}{2} - \mu \right); \frac{5}{2}; \lambda r^2 \right\rbrace \right] \left. {}_1^1\mathrm{F} \left\lbrace \frac{1}{2} \left( \frac{3}{2} - \mu \right); \frac{3}{2}; \lambda r^2 \right\rbrace \right] dr 
\\  
- \frac{{\hbar}^2}{2m_w^*} \int_0^R \mathrm{exp} \left( - \lambda r^2 \right) \mathrm{exp} \left( -2 \beta r \right) \left[ {\lambda}^2 r^4 + 2 \lambda \beta r^3 + \left\lbrace {\beta}^2 - \left( \frac{3}{2} + \mu \right) \right\rbrace r^2 \right. 
\\  
- \left. \left( \frac{1}{2}  + \mu \right) \beta r  - \left( \frac{3}{2} - \mu \right) \right] {\left[ {}_1^1\mathrm{F} \left\lbrace \frac{1}{2} \left( \frac{3}{2} - \mu \right); \frac{3}{2}; \lambda r^2 \right\rbrace \right]}^2 dr + \int_0^R \mathrm{exp} \left( - \lambda r^2 \right) \\
\mathrm{exp} \left( -2 \beta r \right) \left[ \frac{V_0}{R^2}r^4  - \frac{q^2}{4 \pi {\epsilon}_0 {\epsilon}_w} r - \frac{q^2}{4 \pi {\epsilon}_0 {\epsilon}_w R} \left( \frac{1}{{\epsilon}_b} - \frac{1}{{\epsilon}_w} \right) r^2  + \frac{1}{2} \frac{q^2}{4 \pi {\epsilon}_0 {\epsilon}_w} \right.
\\
\left. \sum_{n=0}^{\infty} \frac{\left(n+1\right) \left( {\epsilon}_w - {\epsilon}_b \right) }{n {\epsilon}_w + \left( n+1 \right) {\epsilon}_b} 
\frac{1}{R^{2n+1}} r^{2n+2} \right] {\left[ {}_1^1\mathrm{F} \left\lbrace \frac{1}{2} \left( \frac{3}{2} - \mu \right); \frac{3}{2}; \lambda r^2 \right\rbrace \right]}^2 dr  + R^2 \mathrm{exp} \left( - \lambda R^2 \right)
\\ 
{\left[ {}_1^1\mathrm{F}\left\lbrace \frac{1}{2} \left( \frac{3}{2} - \mu \right); \frac{3}{2}; \lambda R^2 \right\rbrace \right]}^2 \mathrm{exp} \left( 2 \chi R \right) \left[ \left\lbrace -\frac{{\hbar}^2}{2m_b^*} {\left( \chi + \beta \right)}^2 + V_0 \right\rbrace - \frac{q^2}{4 \pi {\epsilon}_0 {\epsilon}_b r} \right. 
\\  
+ \left. \left. \frac{q^2}{8\pi {\epsilon}_0 {\epsilon}_w} \sum_{n=0}^{\infty} \frac{n \left( {\epsilon}_b - {\epsilon}_w \right) }{n {\epsilon}_w + \left( n+1 \right) {\epsilon}_b} \frac{1}{r^{2n+2}}  R^{2n+1} \right]  \int_R^{\infty} \mathrm{exp} \left\lbrace -2 \left( \chi + \beta \right) r \right\rbrace dr \right]
\end{multline}

The binding energy of an on-center hydrogenic donor can be finally estimated as 
\begin{equation}
\label{eq12}
E_b=E_0-E_0^'
\end{equation}
\noindent where $E_0^'$ is the minimum value of $E_0^'\left(\beta\right)$ w.r.t. $\beta$.


\section{Results and discussion}
\label{Result}

\begin{figure}[t]%
	\centering
	\includegraphics*[width=0.75\linewidth]{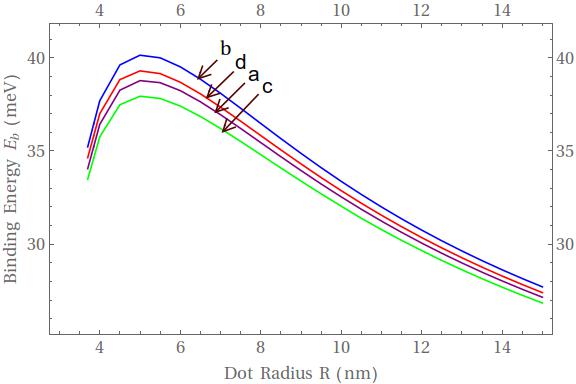}
	\caption{%
		Ground state donor binding energy as a function of the size of $\mathrm{GaAs-{Al}_{0.3}{Ga}_{0.7}As}$ QD: (a) without polarization and self energy, (b) with polarization only, (c) with self energy only and (d) with both the polarization and self energy.}
	\label{fig1}
\end{figure}

\begin{figure}[t]%
	\centering
	\includegraphics*[width=0.75\linewidth]{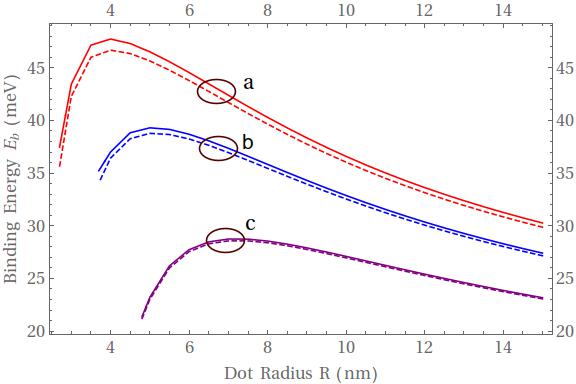}
	\caption{%
		Ground state donor binding energy in the presence (solid) and absence (broken) of polarization and self-energy as a
		function of the size of $\mathrm{GaAs-{Al}_x{Ga}_{1-x}As}$ QD of different potential barriers: (a) 0.348 eV for $x=0.45$, (b) 0.232 eV for $x=0.3$ and (c) 0.116 eV for $x=0.15$.}
	\label{fig2}
\end{figure}

In this communication, $\mathrm{GaAs-Al_{x}Ga_{1-x}As}$ spherical QD with finite parabolic confinement is used to compute the ground state electronic energy, and thereby, the binding energy ($E_b$) associated with the shallow hydrogenic donor. The band gap discontinuity is determined by the $\mathrm{Al}$-composition in the barrier material $\mathrm{Al_x Ga _{1-x} As}$, and is expressed as $\Delta E_g \left(  \mathrm{eV} \right) = 1.247x$, which is distributed 62\% at the conduction band and the rest in the valence band \cite{bib22}. For an electron confined within the dot, the conduction band offset forms the barrier. Material parameters used for computation are taken from ref. \cite{bib22}. As binding energy is the highest for an on-center impurity \cite{bib23}, we have restricted our numerical computations only for such impurity.

\par Figure \ref{fig1} exhibits the ground state binding energies for a donor in $\mathrm{GaAs-}$  $\mathrm{Al_{0.3} Ga_{0.7} As}$ spherical QD as a function of the dot size. To find the individual and finally the overall contribution of polarization and self-polarization charges on impurity binding energy, curves for various situations - in absence and presence of above charges are presented in the figure.

\par The overall binding energy is found to increase due to consideration of both kinds of charges induced at the dot boundary. As the dielectric constant of GaAs dot is larger than that of the $\mathrm{Al_{0.3} Ga_{0.7} As}$ barrier, the ionized donor located at the center of the dot induces polarization charges of the same polarity at the dot-barrier interface \cite{bib17,bib18}. The electron acquires additional potential energy due to interaction with these induced charges. The electron-ionized donor interaction is thus strengthened and donor binding energy increases. Similarly, an electron strictly confined within the dot induces charges of its same polarity at the dot boundary \cite{bib17,bib18}. Interaction of the electron with such self-induced charges being repulsive in nature weakens the overall electron-donor interaction, and reduces the binding energy $E_b$. Thus, self energy partly compensates the effect of polarization energy. In a finite barrier QD, a part of the electron always spills over into the barrier, and the effects of both the polarization and self energy discussed above are reduced to some extent. The overall increase in binding energy is still noticeable. Further, strong spatial confinement of carriers in small dots gives rise to strong interactions. Thus, binding energy increases with decrease in the dot size, and peaks at a certain dot radius. For further reduction in dot size, binding energies in all above cases fall drastically due to lack of effective carrier confinement, and eventually converge.

\par In Fig. \ref{fig2}, results for the model with both the polarization and self-energy contributions are shown for three different $\mathrm{Al}$ concentrations in the embedding $\mathrm{\left(Al,Ga\right)As}$ alloy. A larger value of $x$ implies higher barrier for the finite QD.

\par In the figure, the impurity binding energy is found to be larger in dots with higher barriers. Stronger confinement of carriers in such dots results in higher value of the ground state electronic energy as well as stronger interactions. Thus, for $x=0.45$, value of the binding energy peak is maximum, and it appears for the smallest dot, as expected. On the other hand, binding energies in all three cases tend to converge due to spreading of electron wave functions in too large dots.


\section{Conclusion}
\label{Conclusion}

The overall change in impurity binding energy due to polarization and self-polarization energy is calculated for an on-center impurity within a spherical QD with parabolic confinement. Our calculations indicate an increase of 1.2\% in the maximum binding energy ($E_b$) after inclusion of the polarization and self-energy effects in the $\mathrm{GaAs/ Al_{0.3} Ga_{0.7} As}$ system, almost similar to the variations reported for a square well potential\cite{bib20}. However, for a parabolic potential the binding energy peak appears at 3.9 nm, contrary to the peak at 3.1 nm in the case of the square well potential, implying weaker confinement. Here the influence of the polarization and self-energy effects is strongest for the $\mathrm{GaAs/Al_{0.45} Ga_{0.55} As}$ material combination. A 2.2\% enhancement in binding energy is obtained for nearly 11\% mismatch in static dielectric constant ($\Delta {\epsilon}_r$) between the dot material and the barrier material. In case of larger $\Delta {\epsilon}_r$, a greater amount of induced charges of both kinds will appear at the well-barrier interface and it will have a greater influence on the donor binding energy. The treatment presented in this communication would therefore, be more significant for dot-barrier material combinations with larger dielectric mismatch. 

\section{Acknowledgement}
	Financial support from project UPE-II, UGC, India is gratefully acknowledged. We also acknowledge Dr. Manas Kumar Bose for helpful discussion.


\bibliography{mybibfile}

\begin{thebibliography}{10}
\expandafter\ifx\csname url\endcsname\relax
  \def\url#1{\texttt{#1}}\fi
\expandafter\ifx\csname urlprefix\endcsname\relax\def\urlprefix{URL }\fi
\expandafter\ifx\csname href\endcsname\relax
  \def\href#1#2{#2} \def\path#1{#1}\fi

\bibitem{bib1}
G.~Bastard, Hydrogenic impurity states in a quantum well: A simple model, Phys.
  Rev. B 24 (1981) 4714.
\newblock \href {http://dx.doi.org/10.1103/PhysRevB.24.4714}
  {\path{doi:10.1103/PhysRevB.24.4714}}.

\bibitem{bib2}
G.~Bastard, Hydrogenic impurity states in a quantum well, Surf. Sci. 113 (1982)
  165.
\newblock \href {http://dx.doi.org/10.1016/0039-6028(82)90580-5}
  {\path{doi:10.1016/0039-6028(82)90580-5}}.

\bibitem{bib3}
W.~T. Masselink, Y.~C. Chang, H.~Morkoc, Binding energies of acceptors in
  $\mathrm{GaAs-Al_xGa_{1-x}As}$ quantum wells, Phys. Rev. B 28 (1983) 7373.
\newblock \href {http://dx.doi.org/10.1103/PhysRevB.28.7373}
  {\path{doi:10.1103/PhysRevB.28.7373}}.

\bibitem{bib4}
L.~E. Oliveira, L.~M. Falicov, Energy spectra of donors and acceptors in
  quantum-well structures: Effect of spatially dependent screening, Phys. Rev.
  B 34 (1986) 8676.
\newblock \href {http://dx.doi.org/10.1103/PhysRevB.34.8676}
  {\path{doi:10.1103/PhysRevB.34.8676}}.

\bibitem{bib5}
G.~W. Bryant, Hydrogenic impurity states in quantum-well wires: Shape effects,
  Phys. Rev. B 31 (1985) 7812.
\newblock \href {http://dx.doi.org/10.1103/PhysRevB.31.7812}
  {\path{doi:10.1103/PhysRevB.31.7812}}.

\bibitem{bib6}
J.~W. Brown, H.~N. Spector, Hydrogen impurities in quantum well wires, J. Appl.
  Phys. 59 (1986) 1179.
\newblock \href {http://dx.doi.org/10.1063/1.336555}
  {\path{doi:10.1063/1.336555}}.

\bibitem{bib7}
J.~L. Zhu, J.~J. Xiong, B.~L. Gu, Confined electron and hydrogenic donor states
  in a spherical quantum dot of $\mathrm{GaAs-Ga_{1-x}Al_xAs}$, Phys. Rev. B 41
  (1990) 6001.
\newblock \href {http://dx.doi.org/10.1103/PhysRevB.41.6001}
  {\path{doi:10.1103/PhysRevB.41.6001}}.

\bibitem{bib8}
N.~P. Montenegro, S.~T.~P. Merchancano, Hydrogenic impurities in
  $\mathrm{GaAs-(Ga,Al)As}$ quantum dots, Phys. Rev. B 46 (1992) 9780.
\newblock \href {http://dx.doi.org/10.1103/PhysRevB.46.9780}
  {\path{doi:10.1103/PhysRevB.46.9780}}.

\bibitem{bib9}
C.~Bose, Perturbation calculation of impurity states in spherical quantum dots
  with parabolic confinement, Physica E 4 (1999) 180.
\newblock \href {http://dx.doi.org/10.1016/S1386-9477(99)00010-7}
  {\path{doi:10.1016/S1386-9477(99)00010-7}}.

\bibitem{bib10}
C.~Bose, C.~K. Sarkar, Effect of a parabolic potential on the impurity binding
  energy in spherical quantum dots, Physica B 253 (1998) 238.
\newblock \href {http://dx.doi.org/10.1016/S0921-4526(98)00407-4}
  {\path{doi:10.1016/S0921-4526(98)00407-4}}.

\bibitem{bib11}
J.~L. Movilla, J.~Planelles, Off-centering of hydrogenic impurities in quantum
  dots, Phys. Rev. B 71 (2005) 075319.
\newblock \href {http://dx.doi.org/10.1103/PhysRevB.71.075319}
  {\path{doi:10.1103/PhysRevB.71.075319}}.

\bibitem{bib12}
N.~P. Montenegro, S.~T.~P. Merchancano, A.~Latge, Binding energies and density
  of impurity states in spherical $\mathrm{GaAs-(Ga,Al)As}$ quantum dots, J.
  Appl. Phys. 74 (1993) 7624.
\newblock \href {http://dx.doi.org/10.1063/1.354943}
  {\path{doi:10.1063/1.354943}}.

\bibitem{bib13}
F.~J. Ribeiro, A.~Latge, Impurities in a quantum dot: A comparative study,
  Phys. Rev. B 50 (1994) 4913.
\newblock \href {http://dx.doi.org/10.1103/PhysRevB.50.4913}
  {\path{doi:10.1103/PhysRevB.50.4913}}.

\bibitem{bib14}
J.~L. Zhu, X.~Chen, Spectrum and binding of an off-center donor in a spherical
  quantum dot, Phys. Rev. B 50 (1994) 4497.
\newblock \href {http://dx.doi.org/10.1103/PhysRevB.50.4497}
  {\path{doi:10.1103/PhysRevB.50.4497}}.

\bibitem{bib15}
C.~Bose, Binding energy of impurity states in spherical quantum dots with
  parabolic confinement, J. Appl. Phys. 83 (1998) 3089.
\newblock \href {http://dx.doi.org/10.1063/1.367065}
  {\path{doi:10.1063/1.367065}}.

\bibitem{bib16}
D.~B.~T. Thoai, Hydrogenic donor states in semiconductor microcrystallites,
  Solid State Communication 85 (1993) 39.
\newblock \href {http://dx.doi.org/10.1016/0038-1098(93)90915-A}
  {\path{doi:10.1016/0038-1098(93)90915-A}}.

\bibitem{bib17}
J.~M. Ferreyra, C.~R. Proetto, Strong-confinement approach for impurities in
  quantum dots, Phys. Rev. B 52 (1995) R2309.
\newblock \href {http://dx.doi.org/10.1103/PhysRevB.52.R2309}
  {\path{doi:10.1103/PhysRevB.52.R2309}}.

\bibitem{bib18}
M.~Iwamatsu, K.~Horii, Dielectric confinement effects on the impurity and
  exciton binding energies of silicon dots covered with a silicon dioxide
  layer, Jpn. J. Appl. Phys. 36~(10R) (1997) 6416.

\bibitem{bib20}
M.~K. Bose, K.~Midya, C.~Bose, Effect of polarization and self-energy on the
  ground donor state in the presence of conduction band nonparabolicity in
  $\mathrm{GaAs-(Al,Ga)As}$ spherical quantum dot, J. Appl. Phys. 101 (2007)
  054315.
\newblock \href {http://dx.doi.org/10.1063/1.2511785}
  {\path{doi:10.1063/1.2511785}}.

\bibitem{bib21}
C.~Delerue, M.~Lanoo (Eds.), Nanostructures: Theory and Modeling,
  Springer-Verlag Berlin Heidelberg, 2004.

\bibitem{bib22}
S.~Adachi (Ed.), Physical Properties of III-V Semiconductor Compounds, John
  Wiley \& Sons, 1992.

\bibitem{bib23}
C.~Bose, C.~K. Sarkar, Binding energy of impurity states in spherical
  $\mathrm{GaAs-Ga_{1-x}Al_xAs}$ quantum dots, Phys. Status Solidi B 218 (2000)
  461.
\newblock \href
  {http://dx.doi.org/10.1002/1521-3951(200004)218:2<461::AID-PSSB461>3.0.CO;2-U}
  {\path{doi:10.1002/1521-3951(200004)218:2<461::AID-PSSB461>3.0.CO;2-U}}.

\end{thebibliography}

\end{document}